\documentclass[superscriptaddress,amsmath, nofootinbib]{revtex4}
\usepackage{amsfonts}
\usepackage{graphicx}
\usepackage[latin1]{inputenc}
\usepackage{amsmath}
\usepackage{amssymb}
\usepackage[latin1]{inputenc}
\usepackage{amsmath}
\usepackage{amssymb}
\usepackage{amsfonts}
\usepackage{graphicx}
\usepackage[latin1]{inputenc}
\usepackage{amsmath}
\usepackage{amssymb}

\begin{document}
%\draft

%\preprint{APS/123-QED}

\title{Acceleration of particles in Schwarzschild and Kerr geometries}
\author{Walberto Guzm\'an-Ram\'irez }
\email{wguzman@fisica.ugto.mx} \affiliation{Departamento de Matem\'atica, ICE, Universidade Federal de Juiz de Fora, MG,
Brazil}

\author{Alexei A. Deriglazov }\email{alexei.deriglazov@ufjf.edu.br}
\affiliation{Departamento de Matem\'atica, ICE, Universidade Federal de Juiz de
Fora, MG, Brazil} \affiliation{Department of Physics, Tomsk State University, Lenin Prospekt 36, 634050, Tomsk, Russia}

\date{\today}% It is always \today, today,

\begin{abstract}

The Landau-Lifshitz decomposition of spacetime, or (1+3)\,-split, determines the three-dimensional velocity and
acceleration as measured by static observers. We use these quantities to analyze the geodesic particles in
Schwarzschild and Kerr spacetimes. We show that in both cases there is no room for a positive acceleration (repulsion).
We also compute the escape and terminal speeds. The escape speed in the case of a static black hole coincides with the
Newton result.  For the Kerr spacetime, the escape speed depends on the polar angle, showing that a particle needs less
energy to escape  in the direction close to the polar axis.
The terminal speed at the Schwarzschild  horizon and at the Kerr ergosphere turns out to be equal to the speed of light. 
For a local, stationary observer near to a massive particle in geodesic motion on the equatorial plane of the Kerr spacetime,
the analysis of the velocity reveals that such a particle never reaches the ergosphere.
% for the geodesic motion . 
And, in the case of the Schwarzschild spacetime, we found that the geodesic trajectories reach the horizon perpendicularly.
\end{abstract}

\maketitle %\noindent
%%%%%%{\bf DOI:}
%%%%%%%{\bf PACS numbers:} 11.10.Ef, 03.65.Ca \\
%%%%%%%%{\bf Keywords:}

\section{Introduction}

In General Relativity, the lengths and times measured by observer's rods and clocks do not coincide with the
coordinate diferences between  two spacetime events, but should be computed according to the standard prescription
\cite{bib16}.  One should start from considering of two events that occur at the same spatial point, but separated by the
coordinate time $dx^0$. The time $d\tau$, measured by static observer at this point, is assumed to be $cd\tau=\sqrt{-g_{00}}dx^0$.
Then the expressions for time and length intervals between two events separated by the coordinate differences $dx^\mu$
follow from detailed analysis of the notion of simultaneity, the latter is determined with use of null geodesics. This prescription implies,
in particular, the speed of light as the limiting velocity for massive geodesic particles. The  measurable quantities determine the
physically admissible motions. For instance, the circular geodesic, while exists as
a set of points in the narrow region outside the ergosurface, nevertheless is prohibited in this region because the
velocity of the particle following this geodesic would be above the speed of light \cite{Bardeen_1972, Frolov_2011}. This could be
relevant also for analysis of bounds for non circular orbits of spinless and spinning particles \cite{Bini:2017gj, Zimboras_2018}.  The
coordinate quantities should not be confused with the observable quantities. For instance, using the coordinate
(unrenormalized \cite{Spallicci:2017fdk}) velocity considered by Jaffe and Shapiro \cite{Jaffe}, it
is possible to find the regions in Schwarzschild space where the speed of radially ingoing geodesic particle decreases,
so the gravity in this region looks like a repulsive field. But the repulsion disappears, if one use the  semi
renormalized velocity \cite{Cavalleri}.  Concerning the three-acceleration, second derivative of spatial coordinates
with respect to the coordinate time in the geodesic equation can become positive, which also interpreted sometimes as a
gravitational repulsion. For the detailed discussion of a confusion around the subject, see \cite{Spallicci:2017fdk, AAD_2018_2}.

Almost all the works in this subject are focused in the Schwarzschild spacetime, and in the minor number, in the Kerr
spacetime  \cite{Bini:2017gj}. Here, we are interested in the  general formalism
development  in \cite{bib16, Deriglazov:2017jub, Deriglazov:2015wde}. In such formalism,  the 3-velocity and the
3-acceleration  are defined as 3-dimensional covariant quantities. Besides, from the split of the 4-dimensional
spacetime appears a 3-dimensional metric, which should be used to calculate the norm of 3-vectors and projections
between them. The threading formalism has been used to study the behavior of
the spinless and spinning particle in ultra-relativistic regime \cite{Deriglazov2018, Deriglazov:2017jub,
Deriglazov:2015wde}. Studies of the kinematics and dynamics of probe particles in GR  are important
not only in the theoretical physics but  in astrophysics. Understand the  relation between the physical velocity and
acceleration of a particle, measured by an observed close to the source of gravity, with the system of coordinates used
to represent the spacetime\footnote{The coordinate systems are asymptotically flat, where our telescopes and detectors
are located.} is important in order to understand the mechanism of jet formation in a black hole
\cite{AlZahrani:2013ip, Bini:2017gj, Sadeghi:2013gmf}. Besides, to investigate the dynamics of spinless and
spinning particles close to black holes \cite{Armaza:2016ca,
DPW2, LukesGerakopoulos:2014hq}.

The trajectories of particles in GR are determinated by the geodesic equation, here,  we will work with the equation
for the  geodesic line in the reparametrization-invariant form
\begin{eqnarray}\label{L.6.12}
\frac{d}{d\tau}\left(\frac{\dot x^\mu}{\sqrt{-g_{\mu\nu}\dot x^\mu\dot
x^\nu}}\right) + \Gamma^\mu{}_{\alpha\beta}(g) \frac{\dot x^\alpha \dot x^\beta}{\sqrt{-g_{\mu\nu}\dot x^\mu\dot x^\nu}}=0.
\end{eqnarray}
This form for the geodesic equation could be apparently more complicate, however, it will be useful in the analysis of the
conserved quantities.   Depending on the symmetries of the physical system studied, the metric could admit the Killing
vectors. They generate conserved quantities, i.e. if $\xi^\mu$ is a Killing vector ($\xi_{\mu;\nu} + \xi_{\nu;\mu}=0$),
by direct substitution in (\ref{L.6.12}) we can show that the quantity
\begin{equation}\label{Killing-1}
C=\frac{\xi_\mu \dot x^\mu}{\sqrt{-\dot xg\dot x}} \, ,
\end{equation}
is conserved.

The paper is organized as follows. In Sec. \ref{formalism} we introduce the threading formalism and  define the
3-dimensional velocity and the 3-dimensional acceleration. After to consider the geodesic motion in the 3-acceleration
we compute the longitudinal acceleration, acceleration in the direction of the motion, showing that, for a stationary
metric, the particle can not overcome the speed of the light.
In Sec. \ref{Schwarzschild} we apply the definitions and formulas  of the previous section to the Schwarzschild
spacetime. From the conservation law for the energy, we found the terminal velocity of a particle falling  and  the
escape velocity for this spacetime. In the case of the escape velocity we found that coincides with the
non-relativistic result. Besides, we show that the threading formalism reproduces the results of \cite{Cavalleri}.
In Sec. \ref{Kerrspacetime} we considerer the Kerr spacetime. We introduce the expressions for the 3-velocity and
3-acceleration, and compute the terminal and escape velocities for the Kerr spacetime. In this case, the escape
velocity depends on the azimuthal angle. We analyze the circular orbits in the equatorial plane and the radial motion
in the azimuthal axis.
A general discussion of our results is given in Sec. \ref{conclusions}.

\section{Three-dimensional acceleration and speed of light in general relativity}
\label{formalism}

\subsection{Intervals of time and length and the three-dimensional velocity}

In this section, we briefly  discuss the definitions of covariant three-dimensional velocity and acceleration, that
will be used throughout this paper. For a detailed description, see \cite{bib16, Deriglazov:2017jub, Deriglazov:2015wde}. Consider geodesic
motion of a point particle in the Lorentz manifold with some coordinates $x^\mu$ and a given metric $g_{\mu\nu}$.
Formal definitions of the basic three-dimensional quantities, measured by static observer,  can be obtained by
representing the four-interval in $1+3$ block-diagonal form
\begin{eqnarray}\label{La.2}
-ds^2=&&g_{\mu\nu}dx^\mu dx^\nu    \nonumber\\
=&&-c^2\left[\frac{\sqrt{-g_{00}}}{c}\left(dx^0+\frac{g_{0i}}{g_{00}}dx^i\right)\right]^2+\left(g_{ij}-\frac{g_{0i}g_{0j}}{g_{00}}\right)dx^idx^j.
\end{eqnarray}
This prompts to introduce the infinitesimal time and length intervals as follow:
\begin{eqnarray}\label{La.3.0}
dt_0=\frac{\sqrt{-g_{00}}}{c}\left(dx^0+\frac{g_{0i}}{g_{00}}dx^i\right) = -\frac{g_{0\mu}dx^\mu}{c\sqrt{-g_{00}}} \, ,
\end{eqnarray}
\begin{eqnarray}\label{La.3}
dl_0^2=\left(g_{ij}-\frac{g_{0i}g_{0j}}{g_{00}} \right)dx^idx^j\equiv\gamma_{ij}dx^idx^j \, .
\end{eqnarray}
The last equation defines the three-dimensional metric, $\gamma_{ij}$, which will be used to compute norms of
three-dimensional vectors and projection between them. If $dx^\mu$ represent the coordinate differences between
consequent positions of the particle, its speed should be
\begin{equation}\label{La.3.2}
\qquad v = \frac{dl_0}{dt_0}.
\end{equation}
Using the components $dx^i/dx^0$ of coordinate velocity,  we shall introduce the 3-velocity (velocity for short) ${\bf
v}$, with the components
\begin{eqnarray}\label{La.5}
v^i=\left(\frac{dt_0}{dx^0}\right)^{-1}\frac{dx^i}{dx^0},
\end{eqnarray}
where the conversion factor $(dt_0/dx^0)^{-1}$, is obtained from (\ref{La.3.0})
\begin{eqnarray}\label{La.3.1}
\frac{dt_0}{dx^0}=\frac{\sqrt{-g_{00}}}{c}\left(1+\frac{g_{0i}}{g_{00}}\frac{dx^i}{dx^0} \right).
\end{eqnarray}
Defined in this form, the velocity (\ref{La.5}) is consistent with (\ref{La.3.2}), i.e.,
\begin{equation}\label{La.4}
{\bf v}\gamma{\bf v}=v^i\gamma_{ij}v^j=\left(\frac{dl_0}{dt_0}\right)^2 \, .
\end{equation}
Due to these definitions, a particle with the propagation law $ds^2=0$ has the speed ${\bf v}\gamma{\bf v}=c^2$, and
this is a coordinate-independent statement. This immediately follows from the expression of interval (\ref{La.2}), that
with use of equations (\ref{La.3.0})-(\ref{La.4}) can be rewritten in the form similar to the theory of special
relativity
\begin{eqnarray}\label{La.6}
-ds^2=-c^2dt_0^2+dl_0^2=-c^2dt_0^2\left(1-\frac{{\bf v}\gamma{\bf v}}{c^2}\right).
\end{eqnarray}
Now the square root in (\ref{L.6.12}) can be expressed as
\begin{equation} \label{La.6}
\sqrt{-\dot x g\dot x}= \left(\frac{dt_0}{dx^0}\right)\sqrt{c^2 - {\bf v}\gamma{\bf v}} \, .
\end{equation}
For the latter use, we also introduce the four-dimensional quantity
\begin{eqnarray}\label{La.5.1}
v^\mu=\left(\frac{dt_0}{dx^0}\right)^{-1}\frac{dx^\mu}{dx^0}=\left(\left(\frac{dt_0}{dx^0}\right)^{-1}, ~  {\bf v}\right) \, .
\end{eqnarray}

\subsection{The three-dimensional acceleration.}\label{Def-acceleration}
We now turn to the three-dimensional acceleration \cite{Deriglazov:2017jub}.  Before, let us remark the following:
considering the subgroup of spacial transformations
 \begin{equation}\label{T-3}
x^0=x'{}^0, \quad x^i=x^i(x'{}^j), \quad  \frac{\partial x^i}{\partial x'{}^j}\equiv a^i{}_j(x') \, ,
 \end{equation}
under these, the formalism (\ref{La.3.0})-(\ref{La.6}) remains manifestly covariant, i.e.,
\begin{itemize}
\item  $g_{00}$ and the conversion factor (\ref{La.3.1}) are  scalar functions, $g_{0i}$ and the velocity (\ref{La.5})
are vectors, while $g_{ij}$ and $\gamma_{ij}$ are second rank tensors. \item  $g^{ij}$ is a second rank tensor and,
since $g^{ij}\gamma_{jk}=\delta^{i}{}_k$, the inverse metric of  $\gamma_{ij}$ turns out to be
$(\gamma^{-1})^{ij}=g^{ij}$. \item $\Gamma^i_{00}$ transforms as a  vector and $\Gamma^i_{0j}$ transforms as a  tensor.
These affirmations result of the rule of transformation for the connections applied to  (\ref{T-3}).
%\begin{equation}\label{TG}
%\Gamma'^{\alpha}_{\beta\nu} = \frac{\partial x'^\alpha}{\partial x^\sigma} \frac{\partial x^\lambda}{\partial x'^\beta} \frac{\partial x^\mu}{\partial x'^\nu} %\Gamma^\sigma_{\lambda \mu} + \frac{\partial x'^\alpha}{\partial x^\sigma}\frac{\partial^2 x^\sigma}{\partial x'^\beta \partial x'^\nu} \, ,
%\end{equation}
%
\end{itemize}
It should be desirable to write a 3-dimensional acceleration consistent with the formalism (\ref{La.3.0})-(\ref{La.6}).
In order to do this,  we can start with the Eq. (\ref{La.4}). If we derivate this, in relation of $dx^0$, we obtain
\begin{equation}
\frac{d({\bf v}\gamma{\bf v})}{dx^0} = 2\frac{dv^i}{dx^0}\gamma_{ij}v^j + v^i v^j \frac{\partial \gamma_{ij}}{\partial x^k} \frac{dx^k}{dx^0} \, .
\end{equation}
Whence
\begin{equation}\label{Lb.1}
\left(\frac{dt_0}{dx^0}\right)^{-1} \frac{d({\bf v}\gamma{\bf v})}{dx^0} =  2\left[ \left(\frac{dt_0}{dx^0}\right)^{-1}\frac{dv^i}{dx^0}+ \tilde\Gamma^i{}_{jk}(\gamma)v^jv^k \right] \gamma_{in}v^n \, .
\end{equation}
Where the Christoffel symbols, $\tilde\Gamma^i{}_{jk}(\gamma)$, are constructed using  the
3-dimensional metric $\gamma_{ij}(x^k)$
\begin{eqnarray}\label{La.8.1}
\tilde\Gamma^i{}_{jk}(\gamma)=\frac12\gamma^{ia}(\partial_j\gamma_{ak}+\partial_k\gamma_{aj}-\partial_a\gamma_{jk}).
\end{eqnarray}
The Eq. (\ref{Lb.1}) has the form $\frac{dv^2}{dt_0} =2 {\bf v}\cdot {\bf a}$, which prompts to define the 3-acceleration as
\begin{eqnarray}\label{La.8.2}
a^i= \left(\frac{dt_0}{dx^0}\right)^{-1}\frac{dv^i}{dx^0}+\tilde\Gamma^i{}_{jk}(\gamma)v^jv^k = v^k D_k v^i\, .
\end{eqnarray}
Where we  introduce the 3-dimensional covariant derivative $D_k$ of a vector field $\xi^i( x^k)$
\begin{eqnarray}\label{La.8}
D_k\xi^i=\frac{\partial \xi^i}{\partial x_k}+\tilde\Gamma^i{}_{kj}(\gamma)\xi^j.
\end{eqnarray}
As a consequence, the metric $\gamma$ is covariantly constant, $D_k\gamma_{ij}=0$. As we mentioned, the 3-velocity
(\ref{La.5}) transforms as a vector, so does its covariant derivative, as a consequence, $a^i$ defined in
(\ref{La.8.2}) transforms as is required.
The same procedure can be applied to define the acceleration in the general case when the three-dimensional metric
depends explicitly on the coordinate $x^0$,  $\gamma_{ij}(x^0, x^i)$, however for the purposes of this work it is  not
necessary.

At this point, the  construction of  $a^i$  have been  based only on the requirement of covariance. Now, we will to
consider  the geodesic motion follow by a particle in our definition of acceleration. Using the geodesic equation, Eq.
(\ref{L.6.12}), to express $dv^i/dx^0$ in terms of the connections we obtain that
 \begin{eqnarray}\label{La.11}
a^i=\tilde
M^i{}_j\left[\tilde\Gamma^j{}_{kl}v^kv^l - \Gamma^i_{\mu\nu}v^\mu v^\nu  \right],
\end{eqnarray}
where
\begin{eqnarray}\label{La.13}
\Gamma^i_{\mu\nu}v^\mu v^\nu =\left(\frac{dt}{dx^0}\right)^{-2}\Gamma^i{}_{00} + \Gamma^i{}_{jk}v^jv^k +
2\left(\frac{dt}{dx^0}\right)^{-1}\Gamma^i{}_{0k}v^k \, ,
\end{eqnarray}
carries with the information of the geodesic motion and the matrix $\tilde M^i_{\ j}$ in (\ref{La.11}) is given by
\begin{eqnarray}\label{La.17.1}
\tilde M^i{}_j=\delta^i{}_j-\frac{v^i({\bf v}\gamma)_j}{c^2} \, .
\end{eqnarray}
Since $v^i$ and $\gamma_{ij}$ transform as 3-dimensional tensors, we conclude that $\tilde M$ is a tensor. Let us note
that, on the r.h.s of (\ref{La.11}) neither $\Gamma^i_{jk}$ nor $\tilde\Gamma^i_{jk}$  transform as 3-dimensional
tensors, however, the combination  $(\tilde\Gamma^i_{jk}-\Gamma^i_{jk})$ transforms in the proper way. Thus, the
definition of 3-acceleration given in (\ref{La.8.2}) is consistent, with the formalism development at the begin of this
section, after consider the geodesic motion.

As an application of (\ref{La.11}), let us compute the longitudinal acceleration $({\bf v \cdot a}={\bf v}\gamma{\bf
a})$.
 Contracting (\ref{La.11}) with $({\bf v}\gamma)_i$, and using
 that $({\bf v}\gamma)_i\tilde M^i{}_j=\frac{c^2-{\bf v}\gamma{\bf v}}{c^2}({\bf v}\gamma)_j$,
 we obtain the formula for the longitudinal acceleration
\begin{eqnarray}\label{La.20}
{\bf v}\gamma{\bf a}= \left(1-\frac{{\bf v}\gamma{\bf v}}{c^2}\right)({\bf v}\gamma)_i\left[-\Gamma^i_{\mu\nu}v^\mu
v^\nu+\tilde\Gamma^i{}_{kl}(\gamma)v^kv^l\right] \, .
\end{eqnarray}
This implies that ${\bf v}\gamma{\bf a}\rightarrow 0$ as ${\bf v}\gamma{\bf v}\rightarrow c^2$, so we conclude that
\emph{ the maximum speed of a particle in geodesic motion never overcome $c$ }. Such a property is desirable in a
definition of acceleration in GR.

We could define the 3-acceleration, ${\bf \tilde a}$,  as
\begin{eqnarray}\label{a-nc}
\tilde a^i\equiv  &&\left(\frac{dt_0}{dx^0}\right)^{-1}\frac{dv^i}{dx^0}= a^i - \tilde\Gamma^i{}_{jk}(\gamma)v^jv^k \nonumber \\
=&& M^i{}_j\left[\tilde\Gamma^j{}_{kl}v^kv^l - \Gamma^i_{\mu\nu}v^\mu v^\nu  \right] - \tilde\Gamma^i{}_{jk}(\gamma)v^jv^k \, .
\end{eqnarray}
Such an acceleration is not a 3-vector, however, it will be useful to compare our results with other works. Using
(\ref{La.17.1}), we can write ${\bf \tilde a}$ as
\begin{equation}
\tilde a^i=-\tilde
M^i{}_j \Gamma^i_{\mu\nu}v^\mu v^\nu  - \frac{v^i}{c^2} ({\bf v}\gamma)_j \tilde\Gamma^j{}_{kl}v^kv^l \, .
\end{equation}
The longitudinal acceleration in this case is
\begin{equation}
{\bf v}\gamma{\bf \tilde a}= - \left(1-\frac{{\bf v}\gamma{\bf v}}{c^2}\right)({\bf v}\gamma)_i \Gamma^i_{\mu\nu}v^\mu
v^\nu - \frac{{\bf v}\gamma{\bf v}}{c^2} ({\bf v}\gamma)_j \tilde\Gamma^j{}_{kl}v^kv^l \, .
\end{equation}
Then, the definition (\ref{a-nc}) does not guarantee that the limit velocity of a  massive particle is  the speed of the light.

\section{Schwarzschild spacetime}\label{Schwarzschild}

We shall now apply  the previous formalism to the Schwarzschild solution of the Einstein field equations. First, we enumerate some general
results for a stationary and static metric ($\partial_0 g_{\mu\nu} =0$ and $g_{i0}=0$). In this case, the local interval of time
coincides with the proper time and the 3-dimensional metric is just the space-space part of the 4-dimensional metric
\begin{eqnarray}
dt_0=&&\sqrt{-g_{00}} \ dt \, , \label{S.1} \\
dl_0^2=&&\gamma_{ij}dx^idx^j= g_{ij}dx^idx^j \, . \label{S.2}
\end{eqnarray}
Whence, the conversion factor (\ref{La.3.0}) and the three-velocity become
\begin{eqnarray}
 \left(\frac{dt_0}{dx^0}\right)^{-1}=&& \frac{c}{\sqrt{-g_{00}}} \, , \label{S.1a} \\
 v^i = &&\frac{c}{\sqrt{-g_{00}}}\frac{dx^i}{dx^0} \, .   \label{S.4}
\end{eqnarray}
For the connections we have that
\begin{equation}\label{S.3}
\Gamma^i_{jk} =\tilde \Gamma^i_{jk} (\gamma), \quad  \Gamma^i_{j0} =0, \quad \Gamma^i_{00}=-\frac 12 g^{ik} \frac{\partial g_{00}}{\partial x^k} \, .
\end{equation}
Then, the 3-acceleration  (\ref{La.8.2}) becomes
\begin{equation} \label{S.5}
a^i= - \frac 12 \left( \frac{c^2}{g_{00}} \right) \tilde M^i_{ \  j} g^{jk}\frac{\partial g_{00}}{\partial x^k} \, .
\end{equation}
Finally, the longitudinal acceleration (\ref{La.20}) takes the form
\begin{equation}\label{S.7}
{\bf v}\gamma{\bf a}=-\frac 12 \frac{ \left(c^2-{\bf v}\gamma{\bf v}\right)} {g_{00}} \ {\bf v}\cdot {\bf \nabla } g_{00} \, .
\end{equation}

It will be illustrative to express the Schwarzschild solution in two different sets of coordinates, non-isotropic and isotropic coordinates.

\subsection{Non-isotropic coordinates}

We start by writing the  Schwarzschild metric using \emph{non-isotropic coordinates}
\begin{equation}\label{MS1}
-ds^2  = -\left( 1-\frac{2m}{r}\right) c^2dt^2 + \left( 1 - \frac{2m}{r}\right)^{-1} dr^2 +r^2d\theta^2 + r^2\sin^2\theta d\varphi^2 \, ,
\end{equation}
where $m=MG/c^2$. The basic quantities, (\ref{S.1})-(\ref{S.1a}), for these coordinates are
\begin{eqnarray}
&&dt_0 = \sqrt{1-\frac{2m}{r}} \ dt, \quad \left(\frac{dt_0}{dx^0}\right)^{-1}= \frac{c}{\sqrt{1-\frac{2m}{r}} } \, . \label{S.8a} \\
&&\gamma_{rr} =g_{rr} \equiv \left( 1 - \frac{2m}{r}\right)^{-1}, \quad \gamma_{\theta\theta} = r^2, \quad \gamma_{\phi\phi}= r^2 \sin^2\theta \, ,  \\
&&\gamma_{ij}=0 \quad \mbox{for all other components} \, . \nonumber
\end{eqnarray}
Then,  the 3-velocity (\ref{S.4}) takes the form
\begin{equation}\label{S.8}
{\bf v}=(v^r, v^\theta, v^\phi)=\frac{1}{ \sqrt{1-\frac{2m}{r}} } \left(\frac{dr}{dt}, \frac{d\theta}{dt}, \frac{d\phi}{dt}  \right) \, .
\end{equation}
For the connections, $\Gamma^1_{00}=-\frac 12 (1-\frac{2m}{r})\partial_r g_{00}$, and $\Gamma^2_{00}=\Gamma^3_{00}=0$. With these, the acceleration using  non-isotropic coordinates   is
\begin{equation} \label{S.9}
{\bf a} = (a^r , a^\theta , a^\phi)= -\frac{m}{r^2} \left( c^2 - g_{rr} (v^r)^2, -g_{rr} v^r v^\theta , - g_{rr} v^rv^\phi \right) \, .
\end{equation}
Finally,  the longitudinal acceleration (\ref{S.7}) in this system of coordinates becomes
\begin{eqnarray}\label{S.10}
{\bf v}\gamma{\bf a} &&= -\frac {v^r}{2} \frac{ \left(c^2-{\bf v}\gamma{\bf v}\right)} { ( 1-\frac{2m}{r}) } \frac{\partial }{\partial r} \left( 1-\frac{2m}{r}\right) \nonumber\\
&&= -\frac{m}{r^2} \frac{v^r}{1-\frac{2m}{r}} \left( c^2 - v^2\right) \, .
\end{eqnarray}
For future uses, let us calculate the radial component of the non-covariant acceleration (\ref{a-nc}) in non-isotropic coordinates
\begin{equation}\label{radial-1}
\tilde a^r = -\frac{m}{r^2} \left( c^2 - 2 g_{rr} (v^r)^2\right) + \frac{g_{00}}{r}\left( r^2 (v^\theta)^2 + r^2\sin^2{\theta}(v^\phi)^2 \right) \, .
\end{equation}

\subsection{Isotropic coordinates}
We now write the Schwarzschild solution in   \emph{isotropic coordinates}
\begin{equation}\label{MS-N}
-ds^2= -\left( \frac{\rho - \frac{m}{2}}{\rho + \frac{m}{2}} \right)^2 dt^2+ \left( \frac{\rho + \frac{m}{2}}{\rho} \right)^4\left( d\rho ^2 + \rho^2d\theta^2 + \rho^2\sin^2\theta d\phi^2 \right) \, ,
\end{equation}
which can be deduced from (\ref{MS1}) by the transformation
\begin{equation}\label{MS-N.1}
r=\left(1 + \frac{m}{2\rho} \right)^2 \rho \  \rightarrow \ \rho= \frac 12 \left( r - m + (r^2 - 2mr)^{1/2} \right)\, .
\end{equation}
The local time, and the conversion factor, and the 3-dimensional metric, in these coordinates are
\begin{eqnarray}
&&dt_0 = \frac{\rho -m/2}{\rho + m/2} dt, \quad \left(\frac{dt_0}{dx^0} \right)^{-1} = c \frac{\rho + m/2}{\rho - m/2} \,. \label{SN.1}\\
&&\gamma_{\rho\rho} =g_{\rho\rho} \equiv \left( \frac{\rho + \frac{m}{2}}{\rho} \right)^4, \quad \gamma_{\theta\theta} = \left( \frac{\rho + \frac{m}{2}}{\rho} \right)^4\rho^2, \quad \gamma_{\phi\phi}= \left( \frac{\rho + \frac{m}{2}}{\rho} \right)^4\rho^2 \sin^2\theta \, ,  \\
&&\gamma_{ij}=0 \quad \mbox{for all other components} \, . \nonumber
\end{eqnarray}
Then, for the 3-velocity we obtain
\begin{equation}\label{SN.2}
{\bf v} = \frac{\rho + m/2}{ \rho - m/2} \left( \frac{d\rho}{dt} , \frac{d\theta}{dt} , \frac{d\phi}{dt} \right) \, .
\end{equation}
The 3-acceleration expressed in isotropic coordinates results
\begin{equation}\label{SN.3}
{\bf a} = -\frac{m \ g^{\rho \rho}}{(\rho +\frac m2)(\rho -\frac m2)} \left( c^2 - g_{\rho\rho} (v^\rho)^2, -g_{\rho\rho} v^\rho v^\theta , -g_{\rho\rho} v^\rho v^\phi \right) \, ,
\end{equation}
where $g_{\rho\rho}g^{\rho\rho}=1$. Since $a^i$ is a 3-vector, we can obtain (\ref{SN.3}) from (\ref{S.9}) by using the rule of transformation $a'^{i}=\frac{\partial x'^{i}}{\partial x^j} a^j$, with the transformation (\ref{MS-N.1}) with $\theta'=\theta$ and $\phi'=\phi$. Then, the only non trivial factor of transformation is
\begin{eqnarray}
\frac{\partial \rho}{\partial r}\rvert_{\rho}= \frac{\rho^2}{(\rho +\frac m2)(\rho -\frac m2)} \, .
\end{eqnarray}
Finally, the longitudinal acceleration takes the form
\begin{eqnarray}\label{SN.4}
{\bf v}\gamma {\bf a} &&= -\frac { v^\rho}{2} \frac{ \left(c^2-{\bf v}\gamma{\bf v}\right)} { \left( \frac{\rho - \frac{m}{2}}{\rho + \frac{m}{2}} \right)^2 }\frac{\partial }{\partial r} \left( \frac{\rho - \frac{m}{2}}{\rho + \frac{m}{2}} \right)^2 \nonumber\\
&&= -\frac{m \ v^\rho}{(\rho +\frac m2)(\rho -\frac m2)}  (c^2 - v^2) \, .
\end{eqnarray}

The ideas of Repulsive Gravity  have been studied by  using two quantities,  the velocity and the radial acceleration  \cite{Jaffe, Cavalleri}. In the followings paragraphs   we will to analyze these quantities using the formalism introduced and the relativistic conservation of the energy.

\subsection{Conserved quantities}

The coefficients of the Schwarzschild metric, expressed in both isotropic and non-isotropic coordinates, are independent of $x^0$ and $\phi$, hence $\xi^\mu=(1,0,0,0)$ and $\chi^\mu=(0,0,0,1)$ are Killing vectors. Follows from (\ref{Killing-1}) that the constants of the motion associated with these vectors are
\begin{eqnarray}
-E=&&\frac{ g_{00}}{\sqrt{-\dot x g \dot x}}= \frac{-c \sqrt{-g_{00}}}{\sqrt{c^2-{\bf v}\gamma{\bf v}}}  \, . \label{K-1} \\
L=&&\frac{g_{\phi\phi}\frac{d\phi}{dx^0}}{\sqrt{-\dot x g \dot x}} =\frac{g_{\phi\phi} v^\phi}{\sqrt{c^2-{\bf v}\gamma{\bf v}}} \, , \label{K-1a}
\end{eqnarray}
respectively\footnote{The formula (\ref{K-1}) is valid even if $g_{0i}\ne 0$, we show this result in the next section or can be consulted in \S 88 of \cite{bib16}. }, where we used (\ref{La.6}) and (\ref{S.1a}). In the following paragraphs we apply the conservation law of energy to falling particles.

\subsection{Falling particles}

From (\ref{K-1}) we have
\begin{equation}\label{K-5}
\frac {c^2}{E^2}= \frac{c^2-{\bf v}\gamma{\bf v}}{-g_{00}} \, .
\end{equation}
We can use this constant of motion in order to relate the velocity of a particle in two points of its geodesic trajectory, i.e., $E(r_1)=E(r_2)$.  In particular, if we considerer  one point far away from the origin, where the particle has velocity $v^2_{\infty}$, and another point with spatial coordinates $(r,\theta , \phi)$, with $r > 2m$ ($\rho_0>m/2$ in the case of isotropic coordinates), the equation of conservation (\ref{K-5}) tell us that
\begin{equation}\label{K-5a}
 \frac{c^2-{\bf v}\gamma{\bf v}}{-g_{00}}\rvert_{r\rightarrow \infty} = \frac{c^2-{\bf v}\gamma{\bf v}}{-g_{00}} \rvert_{r>2m}  \, .
\end{equation}
For both cases,  (\ref{MS1}) and (\ref{MS-N}),  $g_{00}\rightarrow -1$ at this limit, then we have that
\begin{eqnarray}\label{K-5b}
 \frac{c^2-{\bf v}\gamma{\bf v}}{-g_{00}} = c^2- v^2_{\infty}  \, .
 \end{eqnarray}
We can use this equation in two directions.
\subsubsection{Terminal velocity}
First, let us considerer a particle which starts its motion at a far distant from the black hole with speed $v^2_\infty$ and reaches some point with coordinates $(r,\theta,\phi)$ ($(\rho, \theta,\phi)$ for isotropic coordinates). The velocity at this point should be
\begin{eqnarray}
\left(\frac{dl_0}{dt_0} \right)^2 = && v_{\infty}^2+\frac{2m}{r}\left( c^2 - v^2_{\infty} \right) \quad \mbox{\emph{for non-isotropic coordinates}} \, , \label{V-local-1}\\
\left(\frac{dl_0}{dt_0} \right)^2 = && \frac{ \frac{2mc^2}{\rho} + v_{\infty}^2(1 - \frac{m}{2\rho})^2}{(1+\frac{m}{2\rho} )^2} \quad \mbox{\emph{for isotropic coordinates}} \, . \label{V-local-2}
\end{eqnarray}
In both cases, the final velocity always increases towards $c^2$ as the particle is approaching to  the
horizon, so we have no repulsive gravity outside of the horizon. These results coincide with the ones reported in \cite{Cavalleri}, this should not be surprising
because our definitions of local intervals of time and length coincide with those of  \cite{Cavalleri}.
\subsubsection{Escape velocity}
As second application the formula (\ref{K-5b}) we can compute the final velocity of a particle which begins its motion at some point $(r,\theta,\phi)$   with velocity ${\bf v}\gamma{\bf v}$ and reach the infinity with velocity $v_{\infty}$, i.e.,
\begin{equation}\label{K-5c}
\frac{v^2_{\infty}}{c^2} =\left( {1-\frac{2m}{r}} \right)^{-1} \left( \frac{{\bf v}\gamma{\bf v}}{c^2} - \frac{2m}{r}  \right) =\frac{\beta^2 -\beta^2_{N} }{1-\beta^2_{N}}\, ,
\end{equation}
 where $ \beta^2_{N} \equiv \frac{2m}{r}=\frac{2MG}{c^2r}$. From (\ref{K-5c}), we conclude that  only particles with initial speed such that $\frac{{\bf v}\gamma{\bf v}}{c^2}> \beta^2_{N}$ can reach the infinity, that is, the escape velocity for the Schwarzschild spacetime coincides with the Newtonian result.

Let us remark that the Eq. (\ref{K-5b}) only relates the magnitudes of the velocity of the particle in two different points along its geodesic trajectory.
% it does not tell us anything about the process from which it  acquired its initial velocity.
%
\subsection{Radial acceleration}
Let us continue with the study of radial motion ($v^\theta=v^\phi=0$), for this case  we have that $v^2= g_{11}(v^1)^2$, and  from (\ref{S.9}) and (\ref{SN.3}) we have
\begin{eqnarray}
a^r=-\frac{m}{r^2}\left( c^2 - v^2 \right) \, , \\
a^\rho = - \frac{mg^{\rho\rho}}{\rho^2- (m/2)^2} \left( c^2 - v^2 \right) \, ,
\end{eqnarray}
using (\ref{K-5b}) we obtain
\begin{eqnarray}
a^r= &&-\frac{m}{r^2}\left(1-\frac{2m}{r}\right)\left[ c^2-v_{\infty}^2 \right] \quad \mbox{\emph{non-isotropic coordinates}} \, ,\label{a-r1}\\
a^\rho= &&-m \frac{\rho - \frac{m}{2}}{\rho + \frac m2} \left( \frac{\rho}{ \rho + \frac m2}\right)^4 \frac{c^2 - v_{\infty}^2}{(\rho+ \frac m2)^2} \quad \mbox{\emph{isotropic coordinates}} \, . \label{a-r2}
\end{eqnarray}
As we can see, the radial acceleration always is negative outside of the horizon,
$r > 2m$ in the case of non-isotropic coordinates and $\rho>m/2$ for isotropic coordinates, then,  there is no repulsion. Even more, in both cases the radial acceleration goes to zero when the particle approaches the horizon.

Next, using (\ref{K-5b}) in the formula of longitudinal acceleration (\ref{S.7}) we obtain
\begin{equation}
{\bf v}\gamma{\bf a}=\frac 12 \left(c^2-v^2_{\infty} \right) \  {\bf v}\cdot {\bf \nabla } g_{00}  \, ,
\end{equation}
whence
\begin{eqnarray}
{\bf v }\gamma{\bf a} =&& -\frac{m}{r^2}v^r\left( c^2 - v_{\infty}^2\right) \quad \mbox{\emph{non-isotropic coordinates}}\, , \label{av-1}\\
{\bf v} \gamma {\bf a}=&& -m v^\rho \left( \frac{\rho - \frac m2}{\rho + \frac m2} \right) \frac{c^2 - v_{\infty}^2}{(\rho + \frac m2)^2} \quad \mbox{\emph{isotropic coordinates}} \, . \label{av-2}
\end{eqnarray}
Here, apparently the longitudinal acceleration expressed in non-isotropic coordinates is different to zero at the horizon, however, from (\ref{V-local-1})
we obtain that $v^r= \sqrt{1-\frac{2m}{r}} \ \left(v_{\infty}^2+\frac{2m}{r}\left( c^2 - v^2_{\infty} \right)\right)^{1/2}$, so, the longitudinal acceleration vanishes  at the horizon in both systems of coordinates.
%
%\newpage

\section{Kerr Spacetime}\label{Kerrspacetime}
We now continue with a more interesting case, when the black hole is no longer static and is  axisymmetric. We
 will consider the Kerr solution expressed in Boyer-Lindquist coordinates
\begin{equation}\label{Kerr}
-ds^2 = g_{00}(dx^0)^2 +g_{0\phi} dx^0d\phi + g_{rr} dr^2 + g_{\theta\theta} d\theta^2 + g_{\phi\phi}d\phi^2 \, ,
\end{equation}
where
\begin{eqnarray}
&&g_{00}=-\left(1-\frac{2mr}{\Sigma}\right), \quad g_{0\phi}=- \frac{2mar\sin^2\theta}{\Sigma}\, , \quad g_{\theta\theta}=\Sigma \, , \\
&&g_{rr}=\frac{\Sigma}{\Delta}, \quad g_{\phi\phi}= \sin^2\theta\left(r^2+a^2+\frac{2ma^2r\sin^2\theta}{\Sigma} \right) \, ,
\end{eqnarray}
with the usual abbreviations
\begin{eqnarray}
\Sigma= r^2+a^2\cos^2\theta,  \quad \Delta=r^2-2mr+a^2 \, ,
\end{eqnarray}
where $m=\frac{MG}{c^2}$ and $a^2\le m^2$ (black hole solution)\footnote{In units with $G=c=1$,  $m$ is the mass of the black hole and $a$ is its angular momentum per unit mass.}. In the Kerr spacetime  (\ref{Kerr}), the outer root of the equation $\Delta=0$, $r=r_+=m+\sqrt{m^2-a^2}$,  defines the event horizon. The outer root of $(\Sigma - 2mr)=0$, $r=R_+\equiv m+\sqrt{m^2-a^2\cos^2\theta}$,  defines the outer boundary of the ergosphere region, $r_+<r<R_+$.

Instead of the coordinate velocity and coordinate acceleration of the particle, we will analyze the velocity and the acceleration that could be measured by local observers. Let us start enumerating the basic quantities of the $1+3$ decomposition. The local interval of time and the conversion factor are
\begin{equation}\label{K.0a}
dt_0 = \frac{\sqrt{-g_{00}}}{c}\left(1+ \frac{g_{0\phi}}{g_{00}} \frac{d\phi}{dx^0} \right) dx^0, \quad \left(\frac{dt_0}{dx^0}\right)^{-1}= -\frac{c\sqrt{-g_{00}}}{g_{00}+g_{0\phi}(\frac{d\phi}{dx^0})} \, .
\end{equation}
The three-dimensional metric resulting is diagonal, with components
\begin{eqnarray} \label{K.2}
\gamma_{rr} = g_{rr}, \quad \gamma_{\theta\theta} = g_{\theta\theta}, \quad \gamma_{\phi\phi}=  g_{\phi\phi} - \frac{(g_{0\phi})^2}{g_{00}} \, .
\end{eqnarray}
Then, the three-dimensional velocity (\ref{La.5}) becomes
\begin{equation}\label{K.1a}
{\bf v} = -\frac{\sqrt{-g_{00}}}{g_{00}+g_{0\phi}(\frac{d\phi}{dx^0})} \left( \frac{dr}{dt} , \frac{d\theta}{dt}, \frac{d\phi}{dt} \right)\, .
\end{equation}
To express the acceleration in a compact way, we define the vector
\begin{equation}
F^i= \tilde \Gamma^i_{mn} v^mv^n -\Gamma^i_{\mu\nu}v^\mu v^\nu \, .
\end{equation}
For the metric (\ref{Kerr}), this vector has components
\begin{eqnarray}
F^r=&&\frac{1}{g_{rr}}\left[ \frac 12 (v^\phi)^2\frac{\partial}{\partial r}\left( \frac{(g_{0\phi})^2}{g_{00}}\right) + v^0v^\phi \frac{\partial g_{0\phi}}{\partial r} +
 \frac 12(v^0)^2 \frac{\partial g_{00}}{\partial r} \right] \, , \label{F-r}\\
F^\theta=&&\frac{1}{g_{\theta\theta}}\left[ \frac 12 (v^\phi)^2\frac{\partial}{\partial \theta}\left( \frac{(g_{0\phi})^2}{g_{00}}\right) + v^0v^\phi \frac{\partial g_{0\phi}}{\partial \theta} +
 \frac 12(v^0)^2 \frac{\partial g_{00}}{\partial \theta} \right] \, ,\label{F-t}\\
F^\phi=&&\frac{1}{(g_{0\phi})^2-g_{00}g_{\phi\phi}}\left[ \left( g_{00} \ {\bf v}\cdot {\bf \nabla} \left( \frac{(g_{0\phi})^2}{g_{00}}\right) - g_{0\phi}{\bf v}\cdot {\bf \nabla}g_{0\phi} \right) v^\phi   \right. \nonumber \\
&&- \left. \left( g_{0\phi} {\bf v}\cdot {\bf \nabla} g_{00} - g_{00} \ {\bf v}\cdot {\bf \nabla} g_{0\phi} \right) v^0  \right] \, . \label{F-p}
\end{eqnarray}
Then, in terms of ${\bf F}$, the components of the acceleration are
\begin{eqnarray}
a^r=&& \left( 1-\frac{(v^r)^2}{c^2}\gamma{rr} \right)F^r -\frac{v^r v^\theta}{c^2}\gamma_{\theta\theta} F^\theta - \frac{v^r v^\phi}{c^2}\gamma_{\phi\phi} F^\phi \, , \\
a^\theta=&& -\frac{v^\theta v^r}{c^2}\gamma_{rr} F^r +\left( 1-\frac{(v^\theta)^2}{c^2}\gamma_{\theta\theta} \right)F^\theta - \frac{v^\theta v^\phi}{c^2}\gamma_{\phi\phi} F^\phi  \, ,\\
a^\phi=&&  - \frac{v^\phi v^r}{c^2}\gamma_{rr} F^r  -\frac{v^\phi v^\theta}{c^2}\gamma_{\theta\theta} F^\theta +\left( 1-\frac{(v^\phi)^2}{c^2}\gamma_{\phi\phi} \right)F^\phi \, .
\end{eqnarray}
Finally, the longitudinal acceleration (\ref{La.20}), we obtain
\begin{equation}\label{la-K}
{\bf v} \gamma {\bf a} = \left( 1 - \frac{{\bf v}\gamma{\bf v}}{c^2} \right) ({\bf v}\gamma{\bf F}) \,.
\end{equation}

From ${\bf a}$ and ${\bf F}$ give above,  we can infer some results about radial motion.
% The conditions for the \emph{radial geodesic  motion} in the Kerr spacetime are not trivial.
If we just take $v^\theta=v^\phi=0$, the components $F^\theta$ and $F^\phi$ do not vanish, therefore, the components of the acceleration $a^\theta$ and $a^\phi$ are different to zero. Then, a local observer should detect a ``force'' in the directions where there is no motion, which is not possible.

The same is true in the equatorial plane, i.e., $\theta=\pi/2$, and $v^\theta=0$. Since in this case, there is no dependence on $\theta$, from (\ref{F-t}), we have that  $F^\theta=0$ and consequently $a^\theta=0$. From the expressions for the remaining components of the  acceleration we can see that it is not possible obtain radial geodesic motion in this plane. If we take $v^\phi=0$, the component of acceleration $a^\phi$ is equal to $F^\phi$, which only vanish for $a=0$ (Schwarzschild spacetime).
At  the end of this section, we will analyze the  motion on the axis of symmetry where the radial motion is possible.

\subsection{Conserved quantities}

As in the Schwarzschild spacetime,  we will use the constant of motion  to show that there is no repulsion in the Kerr spacetime.  First, let us show that (\ref{K-1}) holds for a stationary metric. Such a metric admits the time-like Killing vector  $\xi^\mu=(1,0,0,0)$, then, taking as the parameter the coordinate $x^0$,  Eq. (\ref{Killing-1}) implies that
\begin{eqnarray}\label{K.4}
-E=&&\frac{\xi_\mu \dot x^\mu}{\sqrt{-\dot x g \dot x}} =
\frac{(\frac{dt_0}{dx^0})^{-1}g_{0 \mu}\frac{dx^\mu}{dx^0}} {\sqrt{c^2 - {\bf v}\gamma{\bf v}}} \nonumber \\
= &&- \frac{c\sqrt{-g_{00}}}{ \sqrt{c^2 - {\bf v}\gamma{\bf v}}  }
 \, ,
\end{eqnarray}
where we have used the relation  (\ref{La.6}). Then the relation (\ref{K-1}) is true for a stationary and non-static metric.

% For  the metric (\ref{Kerr}) this formula implies that
%\begin{eqnarray}\label{K.4}
%E
%=&&\frac{\xi_0+\xi_3\frac{dx^3}{dx^0}}{\sqrt{-\dot x g \dot x}} =
%\frac{-B-H\frac{d\phi}{dx^0}}{(\frac{dt_0}{dx^0})\sqrt{c^2 - {\bf v}\gamma{\bf v}}} \nonumber \\
%=  \frac{c\sqrt{B}}{ \sqrt{c^2 - {\bf v}\gamma{\bf v}}  }
 %\, .
%\end{eqnarray}
%
On the other hand, the components of (\ref{Kerr}) do not depend
on  $\phi$, so $\chi^\mu=(0,0,0,1)$ is another Killing vector of the Kerr metric. In this case  the conserved quantity is
\begin{eqnarray}\label{K.5}
L&&=\frac{\chi_0 + \chi_3\frac{dx^3}{dx^0}}{\sqrt{-\dot x g \dot x}} = \frac{(\frac{dt_0}{dx^0})^{-1}(g_{0\phi} + g_{\phi\phi}\frac{d\phi}{dx^0})}{\sqrt{c^2 - {\bf v}\gamma{\bf v}}} %\nonumber \\
%&&=-\frac{c\sqrt{B}}{\sqrt{c^2 - {\bf v}\gamma{\bf v}}}\left(\frac{H-G\frac{d\phi}{dx^0}}{B+H\frac{d\phi}{dx^0}} \right)
\, .
\end{eqnarray}
%
%where we used  (\ref{La.6}) and (\ref{K.0a}).
%
%Again, for our purposes is sufficient  considered only
%From (\ref{K.4}) we have now
%
%\begin{equation}\label{K-10a}
%\frac{c^2}{E^2}= \frac{c^2-{\bf v}\gamma{\bf v}}{B}\, .
%\end{equation}
%
It is possible solve (\ref{K.4}) and (\ref{K.5}) for $d\phi/dx^0$, obtaining
\begin{equation}\label{C.2}
\frac{d\phi}{dx^0}= - \frac{D g_{00} + g_{0\phi} }{D  g_{0\phi}+g_{\phi\phi}} \, ,
\end{equation}
whence
\begin{equation}\label{C.3}
\left(\frac{dt_0}{dx^0}\right)^{-1}= c \sqrt{-g_{00}}\left[ \frac{D g_{0\phi} + g_{\phi\phi}}{(g_{0\phi})^2 - g_{00}g_{\phi\phi}}\right] \, .
\end{equation}
Where $D\equiv L/E$ is the impact parameter.
%The expressions (\ref{C.2}) and (\ref{C.3}) are useful in the analysis of trajectories of falling particles on the equatorial plane.

\subsection{Falling Particles}
As in the case of the Schwarzschild spacetime, we can use  (\ref{K.4})  to relate the velocity of one particle at some
point at infinity, ${\bf v}\gamma{\bf v}=v^2_{\infty}$ (constant), with  its  velocity at some point $(r,\theta,\phi)$,
in this case we are considering points with $r >R_+$, then
\begin{equation}\label{K-10b}
 \frac{c^2-{\bf v}\gamma{\bf v}}{g_{00}}\rvert_{r\rightarrow \infty} = \frac{c^2-{\bf v}\gamma{\bf v}}{g_{00}} \rvert_{r>R_+}  \, .
\end{equation}
 Since $g_{00}\rightarrow -1$ when $r\rightarrow \infty$, we obtain
 \begin{equation}\label{K-11}
 c^2-{\bf v}\gamma{\bf v} = -g_{00}(c^2 - v^2_{\infty}).
 \end{equation}
\subsubsection{Terminal velocity}
As a first application of this formula, let us consider the case when the particle begins the motion at infinity. If
the particle reach the point with spatial coordinates $(r, \theta, \phi)$, the velocity at this point should be
\begin{eqnarray}\label{K.10}
{\bf v}\gamma{\bf v}= \left(\frac{dl_0}{dt_0} \right)^2 = c^2-\left( 1- \frac{2mr}{r^2+a^2\cos^2\theta} \right)\left( c^2 - v^2_{\infty} \right), \quad
 \mbox{at} \quad r \ge R_+\, ,
 \end{eqnarray}
From this equation, we can see that the particle always  increases its speed. The terminal velocity is  $c^2$, when the
particle approaches the ergosurface. Then, we can conclude that there is no repulsion in the Kerr spacetime. In the
equatorial plane ($\theta=\frac{\pi}{2}$) we recover (\ref{V-local-1}).

 Now, we consider the motion of a massive particle on the equatorial plane,
 with $r>R_+$ (in the equatorial plane $R_+=2m$ coincides with the radius of Schwarzschild).
 Let us consider the velocity that could be measured by local observers (\ref{K.1a}).
 Using (\ref{C.2}) and (\ref{C.3}) we can express $v^\phi$ in terms of the impact parameter and the components  of the metric
 %the azimuthal velocity and the radial velocity as
 %
 \begin{eqnarray}
 v^\phi= - c\sqrt{-g_{00}} \left[ \frac{D g_{00} + g_{0\phi} }{(g_{0\phi})^2 - g_{00}g_{\phi\phi}} \right] \, .
 %v^r = -\sqrt{-g_{00}} \left[ \frac{\pi g_{0\phi} - g_{\phi\phi} }{(g_{0\phi})^2 - g_{00}g_{0\phi}} \right] \frac{dr}{dt}\, .
 \end{eqnarray}
 The conservation of energy, Eq. (\ref{K.4}), for the equatorial motion ($v^\theta=0$) becomes
 \begin{equation} \label{E.1}
V_\phi^2 + V^2_r   = c^2 \left(1+\frac{g_{00}}{E^2}\right) \, ,
 \end{equation}
 where we have defined the local azimuthal velocity $V_{\phi}$ and the local radial velocity $V_r$ as
\begin{eqnarray}
V_\phi^2 &&\equiv \gamma_{\phi\phi} (v^\phi)^2 =   \frac{c^2 \left( D g_{00} +
g_{0\phi} \right)^2}{(g_{0\phi})^2 - g_{00}g_{\phi\phi}} \label{E.2} \, , \\
V^2_r &&\equiv \gamma_{rr}(v^r)^2 .
\end{eqnarray}
Then, for the geodesic approaching the ergosurface, Eq. (\ref{E.2}) implies that $V^2_\phi = c^2$, with this, from Eq.
(\ref{E.1}), we obtain that $V^2_r=0$.  This means that an exterior geodesic particle on the equatorial plane never
reach the ergosphere from point of view of a local, stationary observer near to the particle.
This can be contrasted with geodesics
in Schwarzschild space, that have an opposite behavior. In this case the local azimuthal velocity becomes
\begin{eqnarray}\label{E.4}
V_\phi^2 =   c^2 D \frac{(- g_{00})}{g_{\phi\phi}} \label{E.3} \, .
 \end{eqnarray}
For the Schwarzschild solution (\ref{MS1}), we have that $V^2_\phi=0$ at the horizon ($r=2m$), with this, using
(\ref{E.1}),  $V^2_r=c^2$. Then, in the Schwarzschild spacetime,  contrary to the Kerr spacetime, the geodesic
trajectories on the equatorial plane reach the horizon perpendicularly.

\subsubsection{Escape velocity}
Another interesting result  concern to the escape velocity. If the particle begins its motion at some point with spatial coordinates $(r, \theta, \phi)$, with velocity ${\bf v}\gamma{\bf v}$, the final speed (at the infinity) should be
\begin{equation}\label{K.12}
\beta^2_{\infty} =  \frac{ \left(\beta^2 - \beta^2_{K} \right) }{1 - \beta^2_{K}} \, ,
\end{equation}
where $\beta^2={\bf v}\gamma{\bf v}/c^2$, and we define  the  escape velocity in the Kerr spacetime as
\begin{equation}\label{K.13}
\beta^2_{K}= \frac{2mr}{r^2+a^2\cos^2{\theta}} \, .
\end{equation}
From (\ref{K.12}), we conclude that only the particles with initial speed such that $\beta^2_{K}<\frac{{\bf v}\gamma{\bf v}}{c^2}<1 $, can be observer at the infinity.
In this case, the escape velocity only coincides with the Newtonian result if the motion is on the equatorial plane ($\theta=\frac \pi2$), and decrease when the motion begins close at the axis of rotation.   We can express the scape velocity of Kerr space time in terms of the one of the Schwarzschild spacetime
\begin{equation}
\beta^2_{K} = \frac{\beta^2_{N}}{1+\frac{a^2}{r^2}\cos^2\theta} \, .
\end{equation}
From this relation, we can see that $\beta^2_{K} \le \beta^2_{N}$. Then, \emph{out of the equatorial plane, one particle needs  less energy for escape from the Kerr gravitational field that  the Schwarzschild field}.

Let us remark two points, to obtain (\ref{K.12}) and (\ref{K.13}) we are not considering equatorial motion and the computation of velocity  involves the 3-dimensional metric components (\ref{K.2}).

\subsection{Circular orbits}

In this section we analyze azimuthal velocity of circular orbits on the equatorial plane.  The conservation law
(\ref{C.2}), implies $\omega\equiv\frac{d\phi}{dt}=\mbox{const}$. In the other hand, using the radial geodesic equation
for a circular orbit, $r(t)=r=constant$,  in the equatorial plane $\theta=\pi /2$
\begin{equation}\label{or10}
g'_{\phi\phi}\left(\frac{d\phi}{dt}\right)^2+ 2cg'_{0\phi}\left(\frac{d\phi}{dt}\right) +c^2 g'_{00}=0 \, ,
\end{equation}
we can obtain $\omega$ as a function of given $r$. Here the  prime means derivative in relation to $r$. By solving
(\ref{or10}) for $\omega$, we find that the clockwise and counterclockwise particles have different orbital velocities
\begin{eqnarray}\label{or11}
\omega_{1,2}=\pm\frac{c}{g'_{\phi\phi}}\sqrt{(g'_{0\phi})^2-
g'_{00}g'_{\phi\phi}}-\frac{cg'_{0\phi}}{g'_{\phi\phi}}=\pm\frac{c\sqrt{mr^3}}{r^3-m a^2}-\frac{mc a}{r^3-m a^2}
%\equiv\pm\omega_s-\delta\omega.
\end{eqnarray}
When $a=0$, they reduce to the Newton expression, $\omega_{1,2}=\omega_N\equiv
c\sqrt{\frac{m}{r^3}}=\sqrt{\frac{MG}{r^3}}$.   Using (\ref{or11}), we can express $m$ and $a$ through $\omega_1$ and
$\omega_2$
\begin{eqnarray}\label{or12}
MG=r^3\frac{4\omega_1^2\omega_2^2}{(\omega_1-\omega_2)^4}, \qquad a=\frac{c(\omega_1+\omega_2)}{2\omega_1\omega_2}.
\end{eqnarray}
So, measurement of orbital velocities of two particles moving in opposite directions can be used to find the mass and spin of the Kerr body.

It is convenient to rewrite $\omega_{1,2}$ in terms of Newton angular velocity $\omega_N$
%and azimuthal velocities,  and $V_N=r\omega_N\equiv c\sqrt{\frac{\alpha}{2r}}$
\begin{eqnarray}\label{or13}
\omega_{1,2}=\pm\frac{c}{\sqrt{r^3/m}\pm a}=\frac{\omega_N}{\pm1+a\omega_N/c}.
\end{eqnarray}
Due to the dragging effect, value of counterclockwise angular velocity is less the clockwise angular velocity. We point
out that both  are non singular functions outside the horizon. Inserting these expressions into the equation for
azimuthal velocity (\ref{E.2}), we obtain
\begin{eqnarray}\label{or14}
V_{\phi 1, 2}=
%\frac{c\sqrt{\triangle}}{\left|c(r-\alpha)V_N^{-1}+a\right|}=
\frac{\pm c\sqrt{m}\sqrt{\triangle}}{(r-2m)\sqrt{r}\pm a\sqrt{m}}.
\end{eqnarray}
Velocity of a counterclockwise particle $V_{\phi 1}(r)$ is a smooth function outside the ergosurface, $2m<r<\infty$,
with the boundary values $V_{\phi 1}(2m)=c$ and $V_{\phi 1}(\infty)=0$. Besides, $V_{\phi 1}(r_a)=c$ at some point
$r_a>2m$ that obeys the equation $\frac{\sqrt{m}\sqrt{\triangle}}{(r-2m)\sqrt{r}+ a\sqrt{m}}=1$. The latter is
equivalent to Bardeen at all \cite{Bardeen_1972} bound: $r^{3/2}-3mr^{1/2}+2m^{1/2}a=0$. Hence there are no
counterclockwise orbits in the region $2m<r<r_a$ because of velocity of the particle along the orbit would be above the
speed of light.

For the clockwise particle we also have $V_{\phi 2}(2m)=c$ and $V_{\phi 2}(\infty)=0$, but denominator of Eq.
(\ref{or14}) in this case vanishes at the value $r_0>2m$ that obeys the equation $(r-2m)\sqrt{r}- a\sqrt{m}=0$. Then
$V_{\phi 2}$ diverges at this point, and at the value $r_c>r_0$ that obeys the equation
$\frac{\sqrt{m}\sqrt{\triangle}}{(r-2m)\sqrt{r}- a\sqrt{m}}=1$ we have $V_{\phi 2}(r_c)=c$. Hence there are no
clockwise orbits in the region $2m<r<r_c$.

\subsection{Axial Motion}

We end this section considering the motion of a particle on the \emph{ azimuthal axis} ($\theta=0$). First we note that the component $g_{\phi\phi}(r, \theta=0)=0$, so the metric (\ref{Kerr}) becomes singular (non invertible). That is because of the set of coordinates which we are using. In order to avoid any effect due that, let us pass to cartesian coordinates $x'^i=(x,y,z)$ by the usual transformation
\begin{eqnarray}
&&x= r\sin{\theta}\cos{\phi}, \quad y=r\sin{\theta}\sin{\phi} \nonumber\\
&&z=r\cos{\theta} \, ,
\end{eqnarray}
In such a system, the Kerr metric on the azimuthal axis  ($x=y=0$) takes the form
\begin{eqnarray}\label{axis1}
-ds^2=-\left(1-\frac{2mz}{z^2+a^2} \right) (dx^0)^2 + \left(1-\frac{2mz}{z^2+a^2} \right)^ {-1} (dz)^2 +\frac{z^2+a^2}{z^2}\left( (dx)^2 +
(dy)^2\right) \,
\end{eqnarray}
with $z>R_{+}$. Using our formula (\ref{K-11}) we can find  the velocity of a particle which begins its motion at the
infinity, on the azimuthal axis
\begin{eqnarray}\label{axis-v1}
\left(\frac{dl_0}{dt_0} \right)^2 = c^2-\left( 1- \frac{2mz}{z^2+a^2} \right)\left( c^2 - v^2_{\infty} \right), \quad
 \mbox{at} \quad z \ge R_+\, .
 \end{eqnarray}
From this,  the terminal velocity, at $R_+=m+\sqrt{m^2-a^2}$,  is the speed of the light.
%
%Eq. (\ref{axis-v1}) does not show the possibility of acceleration of the axial particles up to almost the speed of
%light. This should be compared with results of the recent work \cite{Bini:2017gj}.
%
The escape velocity (\ref{K.13}) in this case is
\begin{equation}\label{axis-v2}
\beta^2_{K}= \frac{2mz}{z^2+a^2} \, ,
\end{equation}
as is expected from (\ref{K.13}).

Applying the definitions of the threading formalism we obtain the azimuthal acceleration
\begin{equation}\label{az-1}
a^z= - m  \frac{z^2-a^2}{(z^2 + a^2)^2}  \left(c^2- v^2\right) \, ,
\end{equation}
where we set $v^x=v^y=0$. From (\ref{az-1}) we can see that the azimuthal acceleration is always negative, moreover, using the conservation of energy (\ref{K.4}), with $g_{00}$ from (\ref{axis1}),  we can write $a^z$ as
\begin{equation}
a^z= - m  \frac{z^2-a^2}{(z^2 + a^2)^2}\left(\frac{z^2-2mz+a^2}{(z^2+a^2)^2} \right)  \left(c^2- v^2_{\infty}\right) \, ,
\end{equation}
from this, we conclude that the azimuthal acceleration goes to zero when the particle is approaching at the horizon (on the azimuthal axis, the ergosphere  coincides with the horizon).
%
%The Kerr metric (\ref{Kerr}) takes the form
%
%\begin{equation}\label{A-2}
%-ds^2= -B (dx^0)^2 + B^{-1}dr^2 \, .
%\end{equation}
%
%Since $v^\theta=0$ and $\gamma_{\phi\phi}=0$, the  velocity results
%
%\begin{equation}\label{A-3}
%{\bf v}\gamma{\bf v} = A (v^r)^2 \, .
%\end{equation}
%
%The no dependence on $\theta$ and (\ref{A-1}) imply that $F^\theta=F^\phi=0$, with this and the condition or radial motion, $v^\theta=v^\phi=0$, the only no vanishing component of the acceleration is
%
%\begin{equation}\label{A-4}
%a^r= \left(1-\frac{{\bf v}\gamma{\bf v}}{c^2}\right) F^r = - M \left(c^2-{\bf v}\gamma{\bf v} \right) \frac{r^2-a^2}{(r^2 + a^2)^2} \, ,
%\end{equation}
%
%where we used (\ref{A-3}). Using Eq. (\ref{K-11}),  the radial acceleration on the axis of symmetry becomes
%
%\begin{equation}\label{A-5}
%a^r= - M  \left(c^2- v^2_{\infty} \right) B\frac{r^2-a^2}{(r^2 + a^2)^2} \, .
%\end{equation}
%
%Which always is negative and, since $B$ vanish in $R_+$, the radial acceleration vanish towards the particle is approaching to the ergosphere.
%
%The longitudinal acceleration, (\ref{la-K}), in this case becomes
%
%\begin{equation}\label{A-6}
%{\bf v}\gamma{\bf a} = -M  \left(c^2- v^2_{\infty} \right) v^r \frac{r^2-a^2}{(r^2 + a^2)^2} \, .
%\end{equation}
%
%Using Eq. (\ref{A-3}) we can express the radial component of the 3-velocity $v^r$ in terms of the metric components, i.e., $(v^r)^2= B(c^2-B(c^2-v^2_{\infty}))$, which shows that the longitudinal acceleration vanish at $R_+$.

\section{Summary}\label{conclusions}

We have introduced the main features of the threading formalism to split the spacetime, proposed by Landau and Lifshitz \cite{bib16} and extended in \cite{Deriglazov:2017jub, Deriglazov:2015wde}. In section \ref{Def-acceleration} we have defined the 3-acceleration of a particle in this formalism, and we make it consistent with the 4-dimensional geodesic motion, Eq. (\ref{La.11}). We define the longitudinal acceleration, projection of the 3-acceleration along the direction of the motion, Eq. (\ref{La.20}), and we show that the speed of a particle in geodesic motion never overcome the velocity of light.

For the motion of particles in the backgrounds of non-rotating black hole (NRBH) and rotating black hole (RBH),  we calculate the 3-velocity, the 3-acceleration, the velocity ${\bf v}\gamma{\bf v}$ and the longitudinal acceleration, ${\bf v}\gamma{\bf a}$.
In the case of the NRBH, we find that the velocity of a particle approaching  the source always increases, this  for the region outside of the horizon, see Eq. (\ref{V-local-1}) and Eq. (\ref{V-local-2}). Then, there is no  repulsive gravity outside of the horizon. Besides, the particle has as a terminal velocity the speed of the light at the horizon.
In the case of RBH, we found the same behavior, but in this case for the region outside of the ergosphere, i.e.,  there is no repulsive gravity outside of the ergosphere, see Eq. (\ref{K.10}). The terminal velocity in this case is compute at the exterior radius of the ergosphere.
These results come from the relativistic conservation of the energy.
Another result concerns to the possibility that a massive particle enters the ergosphere. For the motion on the equatorial plane we found that due the dragging of the RBH,  the azimuthal velocity $V^2_{\phi}\rightarrow c^2$ and the radial velocity $V^2_r\rightarrow 0$, for geodesics approaching to the ergosphere, see Eq.(\ref{E.1}) and (\ref{E.3}).
Hence, the particle approaching the ergosurface will acquire velocity equal to the orbital velocity of circular motion. This means that geodesic particles on equatorial plane never reach the ergosphere from point of view of the observer located near to the particle. If we considerer the limit case of static black hole ($g_{0i}=0$) we have the contrary effect, i.e.,   $V^2_{\phi}\rightarrow 0$ and $V^2_r\rightarrow c^2$, see Eq. (\ref{E.1}) and (\ref{E.4}).  Then, in the Schwarzschild spacetime the geodesic trajectories on the equatorial plane reach the horizon radially.

Concerning  the acceleration, for the radial geodesic motion in the background of  NRBH, we found that the radial acceleration (the radial component of the 3-acceleration) is always negative outside of the horizon,  furthermore, this vanish as the particle is approaching the horizon, see Eq.(\ref{a-r1}) and Eq. (\ref{a-r2}). The longitudinal acceleration presents the same behavior, Eq. (\ref{av-1}) and Eq. (\ref{av-2}).

%From the expression for the 3-acceleration for  the rotating black hole, Eqs. ()-(), we found that the conditions $v^\theta=v^\phi=0$ in general do not imply  radial motion, i.e., the components of the 3-acceleration in these directions do not vanish. In the equatorial plane, the acceleration in the $\theta$-direction is zero, however the acceleration in the $\phi$-direction is not. The radial motion can be achieve only when the motion occurs  on the axis of symmetry.   In this case, we found that the radial acceleration is always negative and vanish as the particle approach to the ergosphere, see Eq. ().

Besides the terminal velocities, using the relativistic law of  conservation of the energy, we obtain the escape velocities, Eq. (\ref{K-5c}) for the background of NRBH and Eq.(\ref{K.12}) for the RBH. In the case of the NRBH, the escape velocity coincides with the one of the Newtonian gravity.
For the Kerr spacetime, the escape velocity depends on the polar angle, see Eq. (\ref{K.13}), this dependence indicates that the energy necessary to escape from the gravitational field of the RBH is maximum in the equatorial plane (equal to the Newtonian escape velocity) and decrease since the initial point of motion is close to the azimuthal axis. Our result is valid only for initial points outside of the ergosphere. It would be interesting, for theoretical physics as well as astrophysics, find the escape velocity in the case when the motion starts inside of the ergosphere, research in this direction is being constructed and will be reported elsewhere.

Using the local 3-acceleration for the Kerr spacetime we conclude that the radial motion it is not possible unless the motion is along the azimuthal axis.  As in the radial geodesic motion for the NRBH, the axial acceleration is always negative, see Eq. (\ref{az-1}). With this, there is no possibility of a local observer to detect a repulsion along a z-axis. Moreover, the terminal velocity is the speed of the light and the acceleration is zero,  as the particle is approaching at the horizon.

On a number of reasons, it would be of interest to repeat the previous analysis for the more realistic case of a spinning particle. First, the spinning particle represents an exceptional example of intrinsically noncommutative and relativistic invariant theory \cite{Deriglazov:2013}, with the spin-induced noncommutativity that manifest itself already at the Compton scale.  This must be taken into account in a number of semiclassical computations involving the spinning particle in an external  electromagnetic and gravitational fields \cite{Der_Pup_2016, DPW2}. The effects due to noncommutativity of position variables are under considerable interest in the current literature \cite{Kai_2018_1, Blaschke_2018, Git_2015, Dasz_2018, Shokri_2017, Sad_2018, Kai_2018_2, Nandi_2018, Kovacik_2018_1, Kovacik_2018_2, Dubey_2018, Gnat_2018_1, Gnat_2018_2, Guha_2018}, and certainly deserve a detailed study in the relativistic-invariant  context of spin-induced noncommutativity. Second, while the explicit form of equations of motion with account of spin is still under debates \cite{Armaza:2016ca, Kopeikin_2018_1, Kopeikin_2018_2, Gerak_2018_1, Gerak_2018_2}, it seem to be clear that they differ from the geodesic equations. Although it is widely assumed that spin effects are small, in some cases they nevertheless could lead to the qualitatively new effects, see \cite{Der_Wal_2018, Armaza_2016, Mukh_2018, Hamada_2018, Git_2017, Faber_2017, Haroon_2018, Peng_2018, Zhang:2018omr}.

%{\bf Acknowledgements}
\begin{acknowledgments}
WGR: This study was financed in part by the Coordena\c c\~ao de Aperfei\c coamento de Pessoal de N\'ivel Superior -
Brasil (CAPES) -Finance Code 001. The work of AAD has been supported by the Brazilian foundation CNPq (Conselho
Nacional de Desenvolvimento Cient\'ifico e Tecnol\'ogico - Brasil),  and by Tomsk State University Competitiveness
Improvement Program.
\end{acknowledgments}
%\newpage

%\def\refname{Bibliography}

\end{document}